\documentclass[doublecol]{epl2}

\title{Multifractal dimensions for critical random matrix ensembles}
\shorttitle{Multifractal dimensions for critical random matrix ensembles}

\author{J. A. M\'endez-Berm\'udez\inst{1} \and A. Alc\'azar-L\'opez\inst{1} \and Imre Varga\inst{2,3}}
\shortauthor{J. A. M\'endez-Berm\'udez \etal}

\institute{
  \inst{1} Instituto de F\'{\i}sica, Benem\'erita Universidad Aut\'onoma de Puebla,
           Apartado Postal J-48, Puebla 72570, Mexico\\
  \inst{2} Elm\'eleti Fizika Tansz\'ek, Fizikai Int\'ezet, Budapesti M\H uszaki \'es
           Gazdas\'agtudom\'anyi Egyetem, H-1521 Budapest, Hungary\\
  \inst{3} Fachbereich Physik und Wissenschaftliches Zentrum f\"ur Materialwissenschaften,
           Philipps Universit\"at Marburg, D-35032 Marburg, Germany
}

\abstract{
Based on heuristic arguments we conjecture that an intimate
relation exists between the eigenfunction multifractal dimensions
$D_q$ of the eigenstates of critical random matrix ensembles
$D_{q'} \approx qD_q[q'+(q-q')D_q]^{-1}$, $1\le q \le 2$.
We verify this relation by extensive numerical calculations.
We also demonstrate that the level compressibility $\chi$
describing level correlations can be related to $D_q$ in a unified way
as $D_q=(1-\chi)[1+(q-1)\chi]^{-1}$, thus generalizing existing relations
with relevance to the disorder driven Anderson--transition.
}

\begin{document}

\maketitle

\section{Introduction}

It is well--known that the spatial fluctuations of the eigenstates in a
disordered system at the Anderson--transition show multifractal 
characteristics\cite{MJ98,EM08} which has been demonstrated recently
in a series of experiments \cite{expmuf}. Therefore the modeling and
analysis of multifractal states has become of central importance 
producing many interesting results. For this purpose random matrix
models have been invoked and studied recently \cite{BG11,BG11b,ROF11}. 

Since the exact, analytical prediction of the multifractal dimensions 
of the states for the experimentally relevant Anderson--transition in 
$d=3$ or the integer quantum--Hall transition in $d=2$ seems to be out 
of reach, it is desirable to search for heuristic relations in order 
to understand the complexity of the states at criticality. In the 
present paper we propose such heuristic relations that are numerically 
verified using various ensembles of random matrices.

The spatial fluctuations of the  eigenstates can be described by a set of
multifractal dimensions $D_q$ defined by the scaling of the inverse mean
eigenfunction participation numbers with the system size $N$:
\begin{equation}
\left\langle \sum_{i=1}^N |\Psi_i|^{2q} \right\rangle \sim N^{-(q-1)D_q} \ ,
\label{Dq}
\end{equation}
where $\left\langle \cdots \right\rangle$ is the average over some 
eigenvalue window and over random realizations of the matrix. For strongly
localized eigenstates these quantities do not scale with system size, i.e.
$D_q\to 0$ for all $q$, while extended states always feel the entire
system, i.e. $D_q\to d$ for all $q$. Multifractal states, on the other
hand, should be described by the series of the $D_q$, which are a 
nonlinear function of the parameter $q$.

Spectral fluctuations can be characterized in many ways. A usual, often
employed quantity is the level compressibility $\chi$, which is 
extracted from the limiting behavior of the spectral number variance as 
$\Sigma^{(2)}(E) = \left\langle n(E)^2 \right\rangle - \left\langle n(E)
\right\rangle^2 \sim \chi E$, where $n(E)$ is the number of eigenstates in 
an interval of length $E$. The spectral fluctuations in a metallic system
with extended states yield a vanishing compressibility, $\chi\to 0$, while
in a strongly disordered insulating system the levels are uncorrelated, so
they are easily compressible, $\chi=1$. However, for the multifractal states
an intermediate statistics exists, $0<\chi <1$, furthermore the spectral and
eigenstate statistics are supposed to be coupled, which has been pointed
out first in Ref.~\cite{CKL96}.

One of the most important generalized dimensions often used in this context 
is the information dimension $D_1$. It is defined through the scaling of 
the mean eigenfunction entropy with the logarithm of the system size:
\begin{equation}
\left\langle -\sum_{i=1}^N |\Psi_i|^2 \ln |\Psi_i|^2 \right\rangle 
              \sim D_1 \ln N \ .
\label{D1}
\end{equation}
A further, well--known and widely used dimension is called the correlation 
dimension $D_2$, which is extracted from the inverse participation number
from Eq.~(\ref{Dq}) using $q=2$.

In a recent work \cite{BG11} Bogomolny and Giraud have shown that in a 
$d$--dimensional critical system the information dimension $D_1$ and 
the level compressibility $\chi$ are simply related as
\begin{equation}
\chi + D_1/d = 1 \ ,
\label{chiD1}
\end{equation}
furthermore the generalized dimensions $D_q$ can be expressed as
\begin{equation}
\frac{D_q}{d} = \left\{
\begin{array}{ll}
\displaystyle
\frac{\Gamma(q-1/2)}{\sqrt{\pi}\Gamma(q)}(1-\chi) \ , &
1-\chi \ll 1 \\
1-q\chi \ , & \chi \ll 1
\end{array}
\right.  \ .
\label{Dqchith}
\end{equation}
These expressions have been shown to be valid for various critical random 
matrix ensembles in Ref.~\cite{BG11}.

As for the critical, three--dimensional Anderson transition and the
two--dimensional quantum--Hall transition it has been shown earlier
that another relation holds between the level compressibility $\chi$ 
and the correlation dimension $D_2$ \cite{CKL96}:
\begin{equation}
2\chi + D_2/d = 1 \ .
\label{chiD2}
\end{equation}
This relation should obviously hold approximately only since 
$0\leq D_2/d\leq 1$ but $0\leq\chi\leq 1$, leaving the range of
validity for the limit of weak--multifractality. 

In the present work we show a series of relations between various
generalized dimensions, $D_q$ and $D_{q'}$, and the level compressibility
$\chi$ allowing for a generalization that for particular cases
yields Eq.~(\ref{chiD1}) exactly and Eq.~(\ref{chiD2}) in the 
appropriate limit. In order to prove that, numerical simulations
of various critical random matrix ensembles will be used. 
Further implications and more details will be presented elsewhere
\cite{tbp}.

\begin{figure}
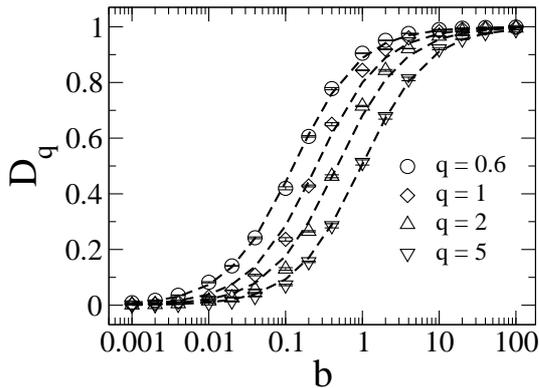

\onefigure[width=7cm]{Fig1.eps}
\caption{$D_q$ as a function of $b$ for the PBRM model
at criticality with $\beta=1$.
The dashed lines are fits of the numerical data with
Eq.~(\ref{Dqofb}).}
\label{Fig1}
\end{figure}
\begin{figure}
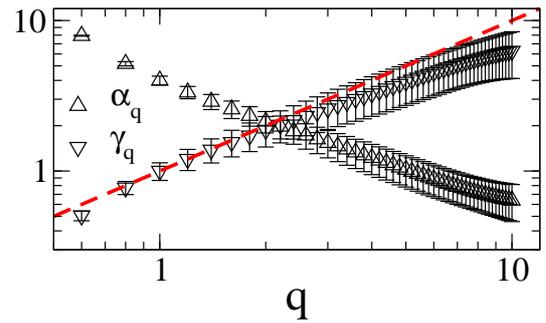

\onefigure[width=7cm]{Fig2.eps}
\caption{$\alpha_q$ and $\gamma_q = \alpha_1/\alpha_q$
as a function of $q$ for the PBRM model at criticality with $\beta=1$. 
The red dashed line equal to $q$ is plotted to guide the eye. 
The error bars are the rms error of the fittings.}
\label{Fig2}
\end{figure}

\section{Model and heuristic relations}

In Ref.~\cite{BG11} Eqs.~(\ref{chiD1}) and (\ref{Dqchith}) were 
shown to be correct numerically for the Power-Law Banded Random Matrix 
(PBRM) model \cite{EM08,MFDQS96,Mirlin00} at criticality. Below we will make 
use of this model to derive our main results.

The PBRM model describes one--dimensional (1d) samples of length $N$ 
with random long-range hoppings. This model is represented by
$N\times N$ real symmetric ($\beta=1$) or complex hermitian ($\beta=2$)
matrices whose elements are statistically independent random variables
drawn from a normal distribution with zero mean and a variance
given by $\langle  |H_{mm}|^2 \rangle =\beta^{-1}$ and
\begin{equation}
   \langle  |H_{mn}|^2 \rangle =
   \frac{1}{2} \frac{1}{1+\left[
   \sin\left( \pi|m-n|/N \right)/(\pi b/N) \right]^{2\mu}} \ ,
\label{PBRMp}
\end{equation}
where $b$ and $\mu$ are parameters. In Eq.~(\ref{PBRMp}) the PBRM model 
is in its periodic version; i.e. the 1d sample is in a ring geometry.
Theoretical considerations \cite{EM08,MFDQS96,Mirlin00,KT00} and detailed
numerical investigations \cite{EM08,EM00b,V03} have verified that the PBRM
model undergoes a transition at $\mu=1$ from localized states for
$\mu >1$ to delocalized states for $\mu < 1$. This transition shows all
the key features of the disorder driven Anderson metal-insulator 
transition\cite{EM08}, including multifractality of eigenfunctions and 
non-trivial spectral statistics. Thus the PBRM model possesses a line of
critical points $b\in (0,\infty)$ in the case of $\mu=1$. In the 
following we will focus on the PBRM model at criticality, $\mu=1$.
By tuning the parameter $b$ the states cross over from the nature of
weak--multifractality ($b\gg 1$) which corresponds to extended--like
or metallic--like states to strong--multifractality ($b\ll 1$) showing
rather localized, i.e. insulator--like states. Meanwhile at the true, 
Anderson transition in $d=3$ or at the integer quantum--Hall transition
in $d=2$, the states belong to the weakly multifractal regime, 
the PBRM model allows for an investigation without
such a limitation. The evolution of the generalized dimensions as a 
function of the parameter $b$ therefore represent this behavior, i.e.
$D_q\to 1$ for $b\gg 1$ and in the other limit of $b\ll 1$ the
multifractal dimensions vanish as $D_q\sim b$ \cite{EM08,Mirlin00}.

Previously, for the PBRM model at criticality with $\beta=1$, we have 
observed that both, $D_1$ and $D_2$ can be approximated simply as
\cite{MV06} $D_1 \approx [1+(\alpha_1 b)^{-1}]^{-1}$ and
$D_2 \approx [1+(\alpha_2 b)^{-1}]^{-1}$ where $\alpha_{1,2}$ are 
fitting constants. This continuous function is a trivial interpolation 
between the limiting cases of low--$b$ and large--$b$ taking the half 
of the harmonic mean of the two as
\begin{equation}
\frac{1}{D_q}=1 + \frac{1}{\alpha_q b}\ ,
\label{interp}
\end{equation}
valid for $q=1$ and $2$. Here we generalize and propose the following 
heuristic expression for a wider range of the parameter $q$
\begin{equation}
D_q \approx \left[ 1+(\alpha_q b)^{-1} \right]^{-1} \ ,
\label{Dqofb}
\end{equation}
as a global fit for the multifractal dimensions $D_q$ of the PBRM model
in both symmetries, $\beta=1$ and $\beta=2$. 
In Fig.~\ref{Fig1} we show fits of Eq.~(\ref{Dqofb}) to numerically
obtained $D_q$ as a function of $b$ for some values of $q$
and in Fig.~\ref{Fig2} we plot the values of $\alpha_q$ extracted
from the fittings.\footnote{
The multifractal dimensions $D_q$ were extracted from the linear fit
of the logarithm of the inverse mean eigenfunction participation numbers
versus the logarithm of $N$, see Eq.~(\ref{Dq}). $D_1$ was extracted 
from the linear fit of the mean eigenfunction entropy versus the logarithm 
of $N$, see Eq.~(\ref{D1}). We used $N=2^n$,
$8\le n\le 13$. The average was performed over $2^{n-3}$ eigenvectors
with eigenvalues around the band center with $2^{16-n}$ realizations of
the random matrices.}
We observe that Eq.~(\ref{Dqofb}) fits reasonably well the
numerical $D_q$ for $q> 1/2$. It is important to stress that
Eq.~(\ref{Dqofb}) reproduces well the $b$-dependencies predicted
analytically \cite{EM08} for the limits $b\ll 1$ and $b\gg 1$.
\begin{figure}[t]
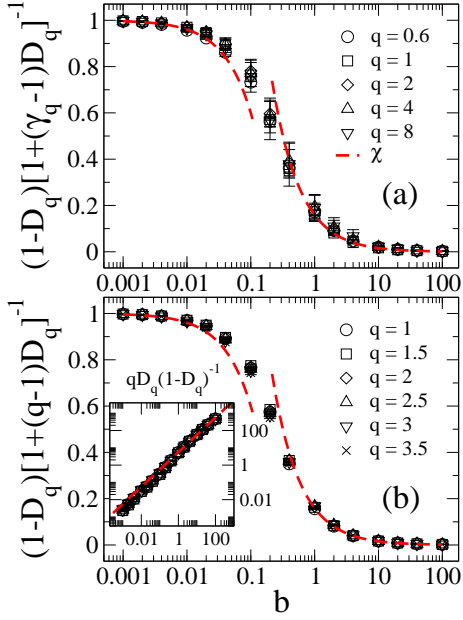

\onefigure[width=6cm]{Fig3.eps}
\caption{(a) $(1-D_q)[1+(\gamma_q-1)D_q]^{-1}$ and (b) $(1-D_q)[1+(q-1)D_q]^{-1}$
[see Eqs.~(\ref{chiofgamma}) and (\ref{chiofDq})] as a function of $b$
for the PBRM model. The red dashed lines are the analytical 
prediction for $\chi$ given in Eq.~(\ref{chithPBRM}).
Inset in (b): $qD_q(1-D_q)^{-1}$ as a function of $b$, see Eq.~(\ref{DqpDqPBRM}).
The red dashed line equal to $\alpha_1b$ is plotted to guide the eye.}
\label{Fig3}
\end{figure}
\begin{figure}[t]
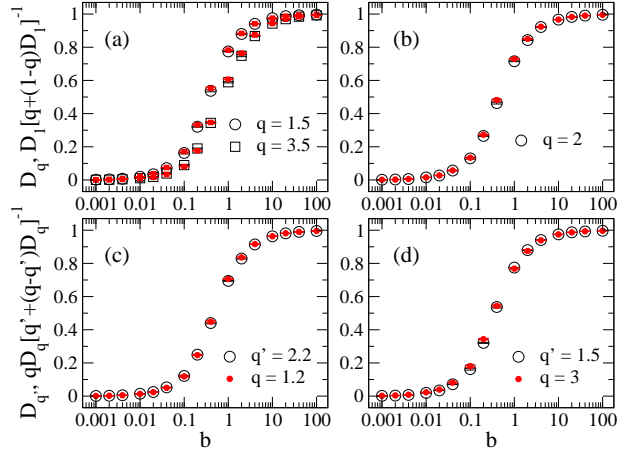

\onefigure[width=8cm]{Fig4.eps}
\caption{(a-b) $D_q$ (open symbols) and $D_1[q+(1-q)D_1]^{-1}$
(full red symbols) [see Eq.~(\ref{DqofD1})] and (c-d) 
$D_{q'}$ (open symbols) and $qD_q[q'+(q-q')D_q]^{-1}$
(full red symbols) [see Eq.~(\ref{DqpofDq})] as a
function of $b$ for the PBRM model.}
\label{Fig4}
\end{figure}
We noticed that by the use of Eq.~(\ref{Dqofb}), Eq.~(\ref{chiD1}) leads to
\begin{equation}
\chi \approx \left( 1+\alpha_1 b \right)^{-1}  \ ,
\label{chiofb}
\end{equation}
which also reproduces well the $b$-dependencies predicted
analytically \cite{BG11,EM08} in the small- and large-$b$ limits:
\begin{equation}
\chi = \left\{
\begin{array}{ll}
1-4b \quad & b \ll 1 \\
(2 \pi b)^{-1} & b \gg 1
\end{array}
\right. \ .
\label{chithPBRM}
\end{equation}
Then, by equating $b$ in Eqs.~(\ref{Dqofb}) and (\ref{chiofb}) we get
\begin{equation}
\chi \approx (1-D_q) \left[ 1+(\gamma_q-1)D_q \right]^{-1} \ ,
\label{chiofgamma}
\end{equation}
with $\gamma_q = \alpha_1/\alpha_q$.
We observed that $\gamma_q \approx q$ in the range $0.8<q<2.5$, 
see Fig.~\ref{Fig2}, so in this range of $q$ values
we can write simplified relations between $\chi$ and $D_q$:
\begin{equation}
\chi \approx \frac{1-D_q}{1+(q-1)D_q} \quad \mbox{and} \quad D_q \approx \frac{1-\chi}{1+(q-1)\chi} \ .
\label{chiofDq}
\label{Dqofchi}
\end{equation}
The expression for $D_q$ in Eq.~(\ref{Dqofchi}) reproduces Eq.~(\ref{Dqchith}) 
exactly for $q=1$ and $q=2$ and approximately for $1<q<2.5$.
Moreover, Eq.~(\ref{Dqofchi}) combined with Eq.~(\ref{chiD1}) allows us
to express any $D_q$ in terms of $D_1$:
\begin{equation}
D_q \approx D_1 \left[ q+(1-q)D_1 \right]^{-1} \ .
\label{DqofD1}
\end{equation}

We also noticed that by equating $\chi$ for different $D_q$'s form
Eq.~(\ref{chiofDq}) we could get recursive relations for them:
\begin{equation}
\frac{q'D_{q'}}{1-D_{q'}} = \frac{qD_q}{1-D_q} \quad \mbox{and} \quad D_{q'} = \frac{qD_q}{q'+(q-q')D_q} \ ,
\label{DqpDq}
\label{DqpofDq}
\end{equation}
which lead to $D_{q+1} = qD_q(1+q-D_q)^{-1}$, when $q'=q+1$.
These expressions also provide a relation between the correlation
dimension and the information dimension or between the
correlation dimension and the compressibility of the spectrum:
\begin{equation}
D_2 = D_1\left( 2-D_1 \right)^{-1} = \left( 1-\chi \right) \left( 1+\chi \right)^{-1} \ .
\label{D2ofD1}
\end{equation}
It is relevant to add that in the weak multifractal regime, 
i.e. when $\chi\to 1$, Eq.~(\ref{D2ofD1}) reproduces the 
relation given in Eq.~(\ref{chiD2}) with $d=1$, reported in \cite{CKL96}.

\begin{figure}[t]
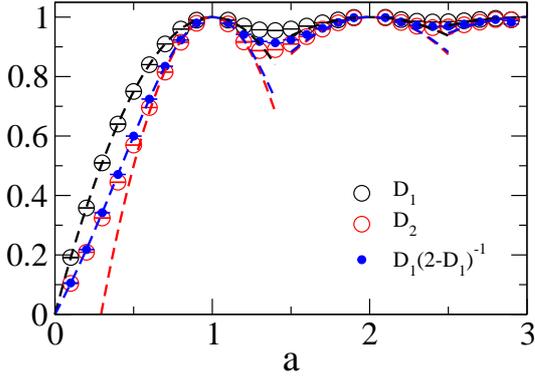

\onefigure[width=7cm]{Fig5.eps}
\caption{$D_1$, $D_2$, and $D_1(2-D_1)^{-1}$ as a function of $a$ for the
RSE (black open, red open, and blue full circles). Black and
red dashed lines are the theoretical predictions for $D_1$ and $D_2$,
respectively, given in Eqs.~(\ref{chiDqRSE}) and (\ref{chiDqRSEk}). The blue
dashed line is the prediction for $D_2$ given by Eq.~(\ref{D2RSE}).}
\label{Fig5}
\end{figure}
\begin{figure}[t]
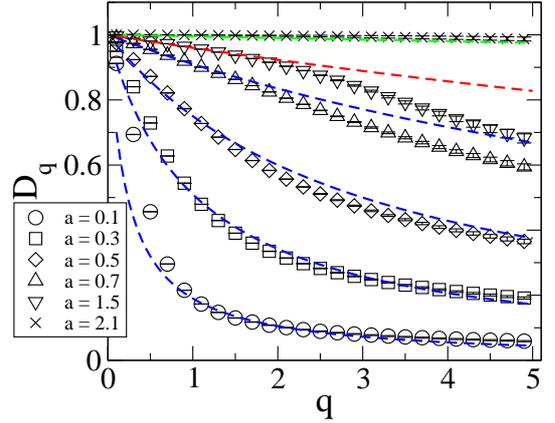

\onefigure[width=7cm]{Fig6.eps}
\caption{$D_q$ as a function of $q$ for the RSE (black open symbols)
for several values of $a$. Blue dashed lines are Eq.~(\ref{DqRSE2}).
The green dashed line is Eq.~(\ref{DqRSE2k}) with $k=2$.
The red dashed line is Eq.~(\ref{DqofD1}) with $D_1(a=1.5)=0.96$.}
\label{Fig6}
\end{figure}

\section{Numerical results for the PBRM model}

Here we verify the expressions (\ref{chiofgamma}-\ref{D2ofD1}) for 
the PBRM model at criticality. Below we concentrate on the case 
$\beta=1$ but we have already validated our results for $\beta=2$.

In Fig.~\ref{Fig3} we plot $(1-D_q)[1+(\gamma_q-1)D_q]^{-1}$ and
$(1-D_q)[1+(q-1)D_q]^{-1}$ as a function of $b$ for several values
of $q$ and observe good correspondence with the analytical prediction
for $\chi$; that is, we verify the validity of Eqs.~(\ref{chiofgamma})
and (\ref{chiofDq}), respectively. 
In the inset of Fig.~\ref{Fig3}(b) we plot $qD_q(1-D_q)^{-1}$ as a 
function of $b$, see Eq.~(\ref{DqpDq}), which for the PBRM model
acquires the simple form
\begin{equation}
qD_q\left( 1-D_q \right)^{-1} \approx q\alpha_q b \approx \alpha_1b \ .
\label{DqpDqPBRM}
\end{equation}

Then, in Fig.~\ref{Fig4} we compare $D_q$ and $D_{q'}$ with 
$D_1[q+(1-q)D_1]^{-1}$ and $qD_q[q'+(q-q')D_q]^{-1}$, respectively,
for several values of $q$; that is, we verify the validity of 
Eqs.~(\ref{DqofD1}) and (\ref{DqpofDq}). Eq.~(\ref{D2ofD1}) is also 
validated in Fig.~\ref{Fig4}(b).

Additionally, in \cite{KOYC10} the duality relation
\begin{equation}
D_2(B) + D_2(B^{-1}) = 1 \ , \quad B\equiv 2^{1/4}\pi b \ ,
\label{duality}
\end{equation}
was shown to be valid (with maximum deviations of $1\%$)
for the PBRM model at criticality. We also want to comment that by the use of
Eq.~(\ref{Dqofb}) we could write $D_2(B) \approx [1+(\delta B)^{-1}]^{-1}$,
with $\delta\equiv \alpha_2/(2^{1/4}\pi)$,
so relation (\ref{duality}) gets the form
\begin{equation}
D_2(B) + D_2(B^{-1}) \approx 1-\frac{B(\delta-1)^2}{B+\delta(B^2+\delta B+1)} \ .
\label{duality2}
\end{equation}
We notice that the quantity $D_2(B) + D_2(B^{-1})$ is very sensitive to 
the value of $\alpha_2$. So, the error in $\alpha_2$ is magnified in the r.h.s. 
of Eq.~(\ref{duality2}).
The maximal deviation from 1 (of $7.3\%$ and $2\%$ for $\beta=1$ and $\beta=2$,
respectively) occurs at $B=1$ where the r.h.s. of Eq.~(\ref{duality2}) acquires 
the form $1-[(\delta-1)/(\delta+1)]^2$. 

\section{Other critical ensembles}

Remember that relations (\ref{chiofDq}-\ref{D2ofD1}) were
obtained form the combination of Eqs.~(\ref{Dqofb}) and (\ref{chiofb}). 
That is, relations (\ref{chiofDq}-\ref{D2ofD1}) are expected to 
work in particular for the PBRM model at criticality. However, 
Eqs.~(\ref{chiofDq}) reproduce Eqs.~(\ref{chiD1}) and (\ref{Dqchith}), 
which were shown to be valid for the PBRM model but also for other 
critical ensembles \cite{BG11}. Then the question is to which extent 
relations (\ref{chiofDq}-\ref{D2ofD1}) are valid for critical ensembles 
different to the PBRM model. So, in the following we verify the 
validity of Eqs.~(\ref{chiofDq}-\ref{D2ofD1}) for other critical 
ensembles.\footnote{
The multifractal dimensions $D_1$ and $D_q$ for those ensembles 
were extracted numerically by the use of the same matrix sizes 
and ensemble realizations as for the PBRM model, if not indicated
otherwise.}

\subsection{The Ruijsenaars-Schneider Ensemble (RSE)}

The RSE proposed in \cite{BGS09} is defined as matrices of the form
\begin{equation}
H_{mn} = \exp(i\Phi_m) \frac{1-\exp(2\pi ia)}{N[1-\exp(2\pi i(m-n+a)/N)]} \ ,
\label{RSE}
\end{equation}
where $1\le m\le n$, $\Phi_m$ are independent random phases distributed
between 0 and $2\pi$, and $a$ is a free parameter independent on $N$.
When $0<a<1$, the compressibility
and the multifractal dimensions take the form \cite{BG11}
\begin{equation}
\chi \sim (a-1)^2 \quad \mbox{and} \quad D_q = 1-q(a-1)^2 \ ;
\label{chiDqRSE}
\end{equation}
while in the vicinity of an integer $k\ge 2$, when $|a-k|\ll1$,
\begin{equation}
\chi \sim (a-k)^2/k^2 \quad \mbox{and} \quad D_q = 1-q(a-k)^2/k^2 \ .
\label{chiDqRSEk}
\end{equation}
As shown in \cite{BG11}, Eqs.~(\ref{chiDqRSE}) and (\ref{chiDqRSEk})
satisfy relation (\ref{chiD1}). Moreover, by direct substitution of
Eqs.~(\ref{chiDqRSE}) [or Eqs.~(\ref{chiDqRSEk})] we verified that
Eqs.~(\ref{chiofDq}-\ref{D2ofD1}) are also satisfied at leading
order in $(a-1)^2$ [$(a-k)^2$].

In Fig.~\ref{Fig5} we plot $D_1$ and $D_2$ as a function of
$a$ for the RSE. Black and red dashed lines are the theoretical
predictions for $D_1$ and $D_2$, respectively, given in 
Eqs.~(\ref{chiDqRSE}) and (\ref{chiDqRSEk}). As it was earlier shown in 
Ref.~\cite{BG11},  the analytical form of $D_q$ given in Eqs.~(\ref{chiDqRSE}) 
and (\ref{chiDqRSEk}) reproduces very well the numerically obtained $D_1$. 
However, we notice that Eq.~(\ref{chiDqRSE}) does not describe well the numerical 
$D_2$, mainly when $a\to 0$. Now, note that by plotting the numerically obtained 
$D_1/(2-D_1)$ we get good agreement with the numerical data for $D_2$, that is 
Eq.~(\ref{D2ofD1}) works well for this model. Then, if we take 
$D_1\approx 1-(a-1)^2$ and $D_1\approx 1-(a-k)^2/k^2$ as theoretical predictions 
for $D_1$ and plug them into Eq.~(\ref{D2ofD1}) we get
\begin{equation}
D_2 \approx \frac{1-(a-1)^2}{1+(a-1)^2} \quad \mbox{and} \quad D_2 \approx \frac{k^2-(a-k)^2}{k^2+(a-k)^2} \ , 
\label{D2RSE}
\end{equation}
for $0<a<1$ and $|a-k|\ll 1$ with $k\ge 2$, respectively; which in fact work much better 
than $D_2\approx 1-2(a-1)^2$ and $D_2\approx 1-2(a-k)^2/k^2$, correspondingly;
see Fig.~\ref{Fig5}.

\begin{figure}[t]
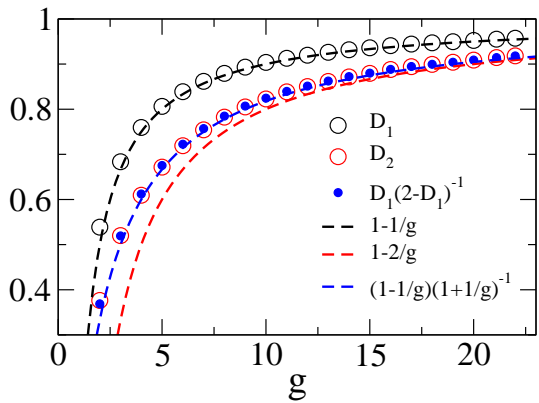

\onefigure[width=7cm]{Fig7.eps}
\caption{$D_1$, $D_2$, and $D_1(2-D_1)^{-1}$ as a function of $g$ for the
IQM model (black open, red open, and blue full circles). Black and
red dashed lines are the theoretical predictions for $D_1$ and $D_2$,
respectively, given in Eqs.~(\ref{chiDqIQM}). The blue dashed line is
the prediction for $D_2$ given by Eq.~(\ref{D2IQM}).}
\label{Fig7}
\end{figure}
\begin{figure}[t]
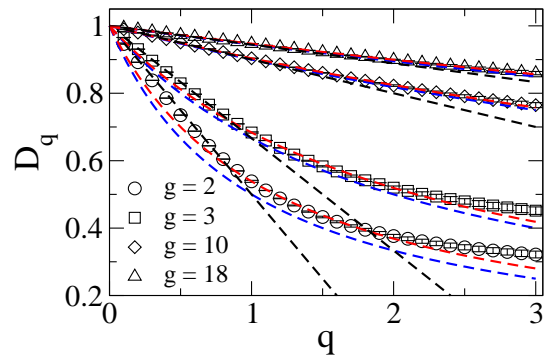

\onefigure[width=7cm]{Fig8.eps}
\caption{$D_q$ as a function of $q$ for the IQM model (black open symbols)
for $g=2$, 3, 10, and 18. Blue dashed lines are Eq.~(\ref{DqIQM2}).
Red dashed lines are Eq.~(\ref{DqofD1}) where the numerical values
$D_1=0.538$, 0.683, 0.904, and 0.947 have been used for $g=2$, 3, 10,
and 18, respectively. Black dashed lines are $D_q$ from Eqs.~(\ref{chiDqIQM}).}
\label{Fig8}
\end{figure}

To get expressions for $D_q$ we substituted $\chi \sim (a-1)^2$ and
$\chi \sim (a-k)^2/k^2$ [or $D_1 \approx 1-(a-1)^2$ and
$D_1 \approx 1-(a-k)^2/k^2$] into Eq.~(\ref{Dqofchi}) [or Eq.~(\ref{DqofD1})],
to get
\begin{equation}
D_q \approx \left[ 1-(a-1)^2 \right] \left[1+(q-1)(a-1)^2 \right]^{-1}
\label{DqRSE2}
\end{equation}
and
\begin{equation}
D_q \approx \left[ k^2-(a-k)^2 \right] \left[k^2+(q-1)(a-k)^2 \right]^{-1} \ .
\label{DqRSE2k}
\end{equation}
In Fig.~\ref{Fig6} we plot $D_q$ as a function of $q$ for the RSE
for several values of $a$. We also plot Eqs.~(\ref{DqRSE2}) and
(\ref{DqRSE2k}) and observe rather good
correspondence with the numerical data mainly in the range $1<q<2$.
Notice that neither Eq.~(\ref{DqRSE2}) nor Eq.~(\ref{DqRSE2k}) can be 
used for $a=1.5$. For that case we substituted the numerically
obtained value of $D_1$ into Eq.~(\ref{DqofD1}) and again observe
good correspondence for $0<q<2$, see the red dashed line in
Fig.~\ref{Fig6}.

\subsection{Intermediate quantum maps}

A variant of the RSE was studied in \cite{MGG08} with the
name of intermediate quantum maps (IQM) model. In this model the parameter
$a$ of the RSE equals $cN/g$ with $cN=\pm 1$ mod $g$, being $g$ the
parameter of the IQM model. For the IQM model the compressibility and the
multifractal dimensions take the form \cite{MGG08}
\begin{equation}
\chi \approx 1/g \quad \mbox{and} \quad D_q \approx 1-q/g \ .
\label{chiDqIQM}
\end{equation}
As for the RSE, here Eqs.~(\ref{chiDqIQM}) satisfy relation
(\ref{chiD1}). Again, by direct substitution of Eqs.~(\ref{chiDqIQM}) 
we verified that Eqs.~(\ref{chiofDq}-\ref{D2ofD1}) are
satisfied at leading order in $1/g$, $g\gg 1$.

We want to mention that in \cite{MGG08} it was shown that
Eq.~(\ref{chiDqIQM}) reproduces well the numerically obtained $D_1$ but
underestimates the numerical $D_2$, in particular for small
$g$, see Fig.~\ref{Fig7}.
Now, notice that by plotting the numerically obtained $D_1/(2-D_1)$ we
nicely reproduce the numerical data for $D_2$, that is Eq.~(\ref{D2ofD1})
works well also for this model. Then, if we take $D_1\approx 1-1/g$ as the
theoretical prediction for $D_1$ and plug it into Eq.~(\ref{D2ofD1}) we get
\begin{equation}
D_2 \approx \left( 1-1/g \right) \left( 1+1/g \right)^{-1} \ ,
\label{D2IQM}
\end{equation}
which in fact works much better than $D_2\approx 1-2/g$ in reproducing the
numerical $D_2$, see Fig.~\ref{Fig7}.
\begin{figure}[t!]
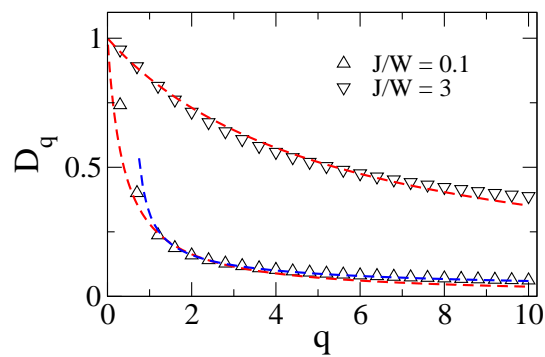

\onefigure[width=7cm]{Fig9.eps}
\caption{$D_q$ as a function of $q$ for the CUE
(taken from \cite{FOR09}).
The blue dashed line is $D_q$ from Eq.~(\ref{chiDqCUE}) for $J/W=0.1$.
Red dashed lines are the prediction for $D_q$ given by
Eq.~(\ref{DqofD1}) using $D_1=0.2805$ and 0.8443 for
$J/W=0.1$ and 3, respectively.}
\label{Fig9}
\end{figure}

To get the expression for $D_q$ we substituted $\chi \approx 1/g$ or
$D_1\approx 1-1/g$ into Eq.~(\ref{Dqofchi}) or (\ref{DqofD1}), respectively,
to get
\begin{equation}
D_q \approx \left( g-1 \right) \left( g+q-1 \right)^{-1} \ .
\label{DqIQM2}
\end{equation}
In Fig.~\ref{Fig8} we plot $D_q$ as a function of $q$ for the IQM model
for some values of $g$. We also plot Eq.~(\ref{DqIQM2}) and observe that it
falls below the numerical data mainly for small $g$. However, by substituting
the numerically obtained values of $D_1$ into Eq.~(\ref{DqofD1}) we get much
better correspondence with the numerical $D_q$, mainly for $1<q<2$.
In Fig.~\ref{Fig8} we also include $D_q$ from Eq.~(\ref{chiDqIQM}). 
We may conclude that
while Eq.~(\ref{chiDqIQM}) reproduces well the numerical $D_q$ for $q<1$,
Eq.~(\ref{DqIQM2}) can serve as the analytical continuation for $q>1$.

\subsection{The critical ultrametric ensemble}

The critical ultrametric ensemble (CUE) proposed in \cite{FOR09}
consists of $2^K \times 2^K$ Hermitian matrices whose matrix elements are
Gaussian random variables with zero mean and variance
\begin{equation}
\left\langle |H_{mm}|^2 \right\rangle = W^2 \ , \quad
\left\langle |H_{mn}|^2 \right\rangle = 2^{2-d_{mn}}J^2 \ ,
\label{CUE}
\end{equation}
where $d_{mn}$ is the ultrametric distance between $m$ and $n$ on the
binary tree with $K$ levels and the root of 1. The parameter in this model
is the ratio $J/W$. For the CUE, when $J/W\ll1$, the compressibility and
the multifractal dimensions have the form \cite{BG11,FOR09}
\begin{equation}
\chi = 1 - \frac{J}{W} \frac{\pi}{\sqrt{2}\ln 2} \quad \mbox{and} \quad
D_q = \frac{J}{W} \frac{\sqrt{\pi}\Gamma(q-1/2)}{\sqrt{2}\ln 2\Gamma(q)} \ .
\label{chiDqCUE}
\end{equation}
Eqs.~(\ref{chiDqCUE}) satisfy relation
(\ref{chiD1}) at first order in $J/W$ \cite{BG11}.
Again, as for the previous critical ensembles, by direct substitution of
Eqs.~(\ref{chiDqCUE}) we verified that
Eqs.~(\ref{chiofDq}-\ref{D2ofD1}) are satisfied at leading order in
$J/W$, for $0.8<q<2.5$; because in this range of $q$ we have that
$\Gamma(q-0.5)/\sqrt{\pi}\Gamma(q)\approx 1/q$.

In Fig.~\ref{Fig9} we show $D_q$ as a function of $q$ for the CUE
for $J/W=0.1$ and 3. The data was taken from \cite{FOR09}.
The blue dashed line is $D_q$ from Eq.~(\ref{chiDqCUE}) for $J/W=0.1$. Notice that
since Eq.~(\ref{chiDqCUE}) is only valid when $J/W\ll 1$ and for
$q\ge 3/4$ one can not use it to predict $D_q$ for $J/W=3$.
However, with Eq.~(\ref{DqofD1}) using as input the numerically
obtained $D_1$ we got good predictions for $D_q$ for small and
large values of $J/W$ and even for values of $q$ smaller than 3/4.
This is shown in Fig.~\ref{Fig9} where we plot
Eq.~(\ref{DqofD1}) (red dashed lines) using $D_1=0.2805$ and 0.8443 for
$J/W=0.1$ and 3, respectively. The values of $D_1$ were obtained by
the interpolation of the $D_q$ data. We observe good correspondence
between Eq.~(\ref{DqofD1}) and the numerical $D_q$ for $0<q<10$.

\section{Conclusions}
In this paper we propose heuristic relations on one hand between the generalized
multifractal dimensions, $D_q$ and $D_{q'}$, for a relatively wide range of the
parameters $q$ and $q'$, and on the other hand between these dimensions and the level 
compressibility $\chi$. As a result we find a general framework embracing an earlier
\cite{CKL96} and a recent one \cite{BG11}. Our proposed relations have been backed 
by numerical simulation on various random matrix ensembles whose eigenstates have 
multifractal properties. These results call for further theoretical as well as 
numerical investigations.

\acknowledgments
The authors are greatly indebted to V. Kravtsov for useful discussions.
This work was partially supported by VIEP-BUAP (Grant No. MEBJ-EXC10-I), 
the Alexander von Humboldt Foundation, and the Hungarian Research Fund (OTKA) 
grants K73361 and K75529.

\end{document}